\documentclass[aps, prl, superscriptaddress, nofootinbib, preprintnumbers,twocolumn]{revtex4}
\usepackage{amsmath}
\usepackage{graphicx}

\newcommand{\lesssim}{\mathrel{\mathpalette\vereq<}}

\newcommand{\chushi}[1]

\begin{document}
 \title{{\bf Dilaton Chiral Perturbation Theory - Determining Mass and Decay Constant of Technidilaton on the Lattice}
 \vspace{5mm}}
\author{Shinya Matsuzaki}\thanks{\tt synya@hken.phys.nagoya-u.ac.jp}
      \affiliation{ Institute for Advanced Research, Nagoya University, Nagoya 464-8602, Japan.}
      \affiliation{ Department of Physics, Nagoya University, Nagoya 464-8602, Japan.}
\author{{Koichi Yamawaki}} \thanks{
      {\tt yamawaki@kmi.nagoya-u.ac.jp}}
      \affiliation{ Kobayashi-Maskawa Institute for the Origin of Particles and 
the Universe (KMI), Nagoya University, Nagoya 464-8602, Japan.}
\date{\today}

\begin{abstract}
We propose a scale-invariant chiral perturbation theory of  the pseudo-Nambu-Goldstone bosons
of chiral symmetry (pion $\pi$) as well as the scale symmetry (dilaton $\phi$) for 
large $N_f$ QCD. 
The resultant dilaton mass $M_\phi$ reads 
$M_\phi^2= m_\phi^2 + \frac{1}{4}(3-\gamma_m)(1+\gamma_m) \cdot (\frac{2 N_f F_\pi^2}{F_\phi^2}) \cdot m_\pi^2$
+ (chiral log corrections), where  
$m_\phi, m_\pi, \gamma_m,  F_\pi$ and $F_{\phi}$ are the dilaton mass in  the  chiral limit, 
the pion mass, the mass anomalous dimension, and the  decay constants  of $\pi$ and $\phi$, respectively. 
The chiral extrapolation of the lattice data, when plotted as 
$M_\phi^2$ vs $m_\pi^2$, 
then simultaneously determines ($m_\phi$, $F_\phi$) of the technidilaton in 
walking technicolor with $\gamma_m \simeq 1$. 
The chiral logarithmic corrections are explicitly given. 

\end{abstract}
\maketitle

Since the Higgs boson was discovered at the LHC~\cite{Aad:2012tfa}, the next stage of particle physics will be 
to elucidate the dynamical origin of the Higgs boson, whose mass and coupling are free parameters within the Standard Model.  
One theory beyond the standard model is Walking Technicolor, which,  based on the approximately scale-invariant gauge dynamics, 
predicted a large anomalous dimension $\gamma_m\simeq 1$ and a pseudo Nambu-Goldstone (NG) boson  of the approximate scale invariance 
(``Technidilaton") as a light composite Higgs boson~\cite{Yamawaki:1985zg}. 
The technidilaton  was actually shown to be consistent with current LHC data for the Higgs~\cite{Matsuzaki:2012mk,Matsuzaki:2012xx}.

A strongly coupled dynamics, walking technicolor would need fully nonperturbative calculations 
in order to make reliable estimates of the properties of the technidilaton and other composite particles to be compared with the upcoming high statistics data at LHC. 
There has been much work on the lattice in search for  
walking technicolor~\cite{Kuti:2014epa}. 
Among others, the LatKMI Collaboration~\cite{Aoki:2013zsa} observed a flavor-singlet scalar meson lighter than 
the ``pion" (corresponding to the NG boson in the chirally broken phase) in $N_f=12$ QCD --- 
a theory shown~\cite{Aoki:2012eq} to be consistent with the chirally unbroken (conformal) phase on the same lattice setting. 
Such a light scalar might be a bound state generated only in the presence of the explicit fermion mass $m_f$ in the conformal phase. 
Still, it gives a good hint for the technidilaton signature in the walking theory, which should have a similar conformal dynamics, 
with the role of $m_f$ instead played 
by the dynamical mass of the fermion generated by 
spontaneous chiral symmetry breaking.

Amazingly, LatKMI Collaboration also observed indications of  a light flavor-singlet scalar with comparable mass to the pion 
in  $N_f=8$ QCD~\cite{Aoki:2013qxa} --- a theory shown~\cite{Aoki:2013xza} to  be  
walking, having both signals of spontaneous chiral symmetry breaking and a remnant of conformality. 
This should be a candidate for the technidilaton as a light composite Higgs boson in walking technicolor.

However, walking technicolor makes sense only for vanishing fermion mass, $m_f\equiv 0$, and hence the techinidilaton mass should be determined in the chiral limit. 
We would need an extrapolation formula for the dilaton mass in the same sense as  the usual chiral perturbation theory (ChPT)~\cite{Gasser:1983yg} 
for the lattice data measured at nonzero $m_f$ to be extrapolated to the chiral limit.

In this article we propose a  scale-invariant ChPT (sChPT) for the use of chiral extrapolation of the lattice data on 
the dilaton and the pion in the presence of explicit mass of the fermion $m_f$. 
It is a scale-invariant generalization of the usual ChPT~\cite{Gasser:1983yg},  
based on the nonlinear realization of chiral symmetry in a way to realize the symmetry structure of 
the underlying walking gauge theory.

The theory consists of the pseudo-NG bosons of the chiral symmetry (pion $\pi$, with mass $m_\pi$) 
as well as the scale symmetry (dilaton $\phi$, with mass $M_\phi$), where both symmetries are broken 
spontaneously by the fermion-pair condensate, and also explicitly by both the fermion mass $m_f$ and the nonperturbative scale anomaly
(induced by the same fermion-pair condensate)~\cite{Miransky:1989qc}. 
We obtain a tree-level formula in  $M_\phi^2$ vs $m_\pi^2$, Eq.(\ref{mphi:p2}), which can be plotted linearly in such a way that the intercept determines 
the chiral limit dilaton mass $m_\phi^2$ (technidilaton mass),  
while its slope gives 
the technidilaton decay constant $F_\phi$, defined as $\langle 0| D^\mu(0)|\phi(q)\rangle =-i F_\phi q^\mu$, and hence 
$\langle 0| \partial_\mu D^\mu(0)|\phi(q)\rangle = -F_\phi M_\phi^2$,
where $D^\mu(x)$ is the dilatation current.   
Based on the sChPT we also explicitly calculate one-loop corrections of the chiral logarithm, Eq.(\ref{mphi:ChiralLog}), 
which turn out to be negligibly small 
in 
current lattice simulations.

Let us start with the chiral/scale Ward-Takahashi (WT) identities for the axialvector ($J^{a \mu}_5$)/dilatation ($D^\mu$) currents  
in the underlying walking gauge theory with 
$N_f$-fermion fields ($\psi$):  
\begin{eqnarray} 
\theta_\mu^\mu=\partial_\mu D^\mu 
&=& \frac{\beta_{\rm NP}(\alpha)} {4\alpha}  G_{\mu\nu}^2   
+ (1+\gamma_m) N_f m_f \bar{\psi}  
\psi  
\,, \nonumber \\  
\partial_\mu J^{a \mu}_5 &=& 2 m_f
\bar{\psi} 
 i \gamma_5 
T^a
\psi   
\,, \label{WI}
\end{eqnarray} 
where $T^a$ $(a=1,\cdots, N_f^2-1)$ are the $SU(N_f)$ generators, and 
$\beta_{\rm NP}(\alpha)$ is the {\it nonperturbative beta function} for the {\it nonperturbative running}~\cite{Miransky:1984ef} of 
the gauge coupling $\alpha$ due to the mass scale $\Lambda_\chi$ 
dynamically generated by the spontaneous breaking of the chiral and scale symmetries through the condensate
 $\langle (\bar \psi \psi)_{\mu=\Lambda_\chi} \rangle\sim -\Lambda_\chi^3$. 
 $\frac{\beta_{\rm NP}(\alpha)} {4\alpha}  G_{\mu\nu}^2$ 
 is the {\it nonperturbative trace (scale) anomaly}~\cite{Miransky:1989qc,footnote1} 
defined as a part associated with the nonpertubative running and is induced solely 
by the chiral condensate with the scale 
$\Lambda_\chi$: $\langle \frac{\beta_{\rm NP}(\alpha)} {4\alpha}  G_{\mu\nu}^2 \rangle|_{m_f=0} \sim -\Lambda_\chi^4$.

We now formulate the sChPT so as to reproduce these WT identities. 
The building blocks $\varphi(x)$ to construct the sChPT are: 
$  \varphi (x) = \{ U(x), \chi(x), {\cal M}(x), S(x) \} $.  
$U(x)=e^{2i \pi(x)/F_\pi}$, $\pi \equiv \pi^a T^a$,  is the usual chiral field with the pion decay constant $F_\pi$, and $\chi(x)=e^{\phi(x)/F_\phi}$ with the dilaton field $\phi(x)$ 
and the decay constant $F_\phi$. 
${\cal M}(x)$ and $S(x)$ are spurion fields introduced so as to incorporate explicit breaking effects of 
the chiral and scale symmetry, respectively.  Under the chiral $SU(N_f)_L \times SU(N_f)_R$ symmetry, these building blocks transform as 
$   U(x) \to g_L \cdot U(x) \cdot g_R^\dag $,  
$  {\cal M} (x) \to g_L \cdot {\cal M}(x) \cdot g_R^\dag $, 
$  \chi(x) \to \chi(x) $ and 
$  S(x) \to S(x) $ with $g_{L,R} \in SU(N_f)_{L,R}$. 
Under the scale symmetry they are infinitesimally transformed as  
$  \delta U(x) = x_\nu \partial^\nu U(x) $,  
$  \delta {\cal M}(x) = x_\nu \partial^\nu {\cal M}(x) $,     
$  \delta \chi(x) =  (1 + x_\nu \partial^\nu) \chi(x)  $ and 
$  \delta S(x) = (1 + x_\nu \partial^\nu) S(x) $, with  
scale dimensions  $d_U=d_{\cal M}=0, d_\chi=d_S=1$.  
The rule of chiral-order counting~\cite{Gasser:1983yg} 
is thus determined consistently with both the scale and chiral symmetries: 
$ U  \sim  \chi  \sim  S \sim {\cal O}(p^0)$, 
$ {\cal M}  \sim  m_f \sim 
{\cal O}(p^2) $,  
$  \partial_\mu  \sim  m_\pi \sim M_\phi \sim {\cal O}(p)$, 
where $m_\pi$ and $M_\phi$ are 
pion and dilaton masses arising from the vacuum expectation values of 
the spurion fields ${\cal M}$ and $S$, $\langle {\cal M} \rangle= m_\pi^2 \times {\bf 1}_{N_f \times N_f}$,  and $\langle S \rangle=1$.

We shall first consider the chiral limit $m_f \to 0$.   
To the leading order ${\cal O}(p^2)$ of sChPT, 
the chiral Lagrangian for the scale-invariant action is uniquely determined as~\cite{footnoteone}:  
\begin{eqnarray} 
 {\cal L}_{(2)}^{\rm inv}
 = \frac{F_\phi^2}{2} (\partial_\mu \chi)^2 
+ \frac{F_\pi^2}{4} \chi^2 {\rm tr}[\partial_\mu U^\dag \partial^\mu U] 
\,.\label{inv:part}
\end{eqnarray}
As noted above, 
even in the chiral limit,
the scale symmetry is explicitly broken by the dynamical generation of the fermion mass itself 
in the underlying walking gauge theory 
(``hard-scale anomaly", or scale violation by the marginal operator) characteristic to the conformal phase transition~\cite{Miransky:1996pd}. 
Hence we have
$4 E= \langle \theta_\mu^\mu\rangle
_{m_f = 0} = \frac{F_\phi}{d_\theta } \langle 0| \theta_\mu^\mu  |\phi \rangle
_{m_f = 0}  
= \frac{F_\phi}{4} \langle 0| \partial_\mu D^\mu |\phi \rangle
_{m_f = 0} =- \frac{F_\phi^2 m_\phi^2}{4} 
 < 0$ 
(Partially Conserved Dilatation Current (PCDC) relation)~\cite{addfootnote}, 
where $m_\phi$ denotes the chiral-limit dilaton mass and we understand 
 that the scale dimension of $\theta_\mu^\mu$ is equal to the canonical dimension,  $d_{\theta}=4$, for $m_f=0$.

We may incorporate the corresponding explicit breaking terms, involving the spurion field $S$ to make the action formally 
scale-invariant~\cite{Matsuzaki:2012vc}:  
\begin{eqnarray} 
{\cal L}^{S}_{(2){\rm hard}} 
=
-\frac{F_\phi^2}{4} m_\phi^2 \chi^4 \left( \log \frac{\chi}{S} - \frac{1}{4}\right)
\,. \label{hard:part}
\end{eqnarray} 
This is a unique form having scale dimension four, which 
correctly reproduces the underlying nonperturbative scale anomaly 
$\frac{\beta_{\rm NP}(\alpha)}{4\alpha} \langle G_{\mu\nu}^2 \rangle$  
in the scale WT identity, Eq.(\ref{WI}), in the chiral limit $m_f\to 0$.
In fact, when $\langle S \rangle =1$, 
non-invariant term arises from $\log \chi$ to yield the scale anomaly 
$\langle \theta_\mu^\mu\rangle =\langle \partial_\mu D^\mu \rangle =\langle \delta {\cal L}^{S}_{(2){\rm hard}} \rangle= -F_\phi^2 m_\phi^2 \langle \chi^4\rangle /4$, 
in accord with the PCDC relation. The last factor $-1/4$ yields a correct vacuum energy 
$E=\langle -{\cal L}^{S}_{(2){\rm hard}} \rangle =-F_\phi^2 m_\phi^2/16= \langle \theta_\mu^\mu\rangle/4$.

 As was discussed in Ref.~\cite{Leung:1989hw},  
the explicit breaking terms due to the 
fermion current mass $m_f$ may also be introduced so as to reproduce the chiral WT identity in Eq.(\ref{WI}):   
\begin{eqnarray} 
 {\cal L}^{S}_{(2){\rm soft}} 
 &=& 
\frac{F_\pi^2}{4} \left( \frac{\chi}{S} \right)^{3-\gamma_m} \cdot S^4 {\rm tr}[ {\cal M}^\dag U + U^\dag {\cal M}] 
\nonumber \\ 
&& 
- \frac{(3-\gamma_m) F_\pi^2}{8}  \chi^4 \cdot  \left( N_f{\rm tr}[{\cal M}^\dag {\cal M}] \right)^{1/2}
\,.   
\label{soft:part}
\end{eqnarray} 
 The factor $(3-\gamma_m)$ in the first line 
reflects  the full dimension of the fermion bilinear operator $\bar{\psi}\psi$ in the underlying gauge theory. 
The scale-invariant term in line two, having no contributions to $\theta_\mu^\mu$, was introduced 
in the case without 
the hard-scale anomaly term ${\cal L}^{S}_{(2){\rm hard}}$~\cite{Leung:1989hw,footnote2} 
in order to stabilize the dilaton potential  so as to make the otherwise tachyonic dilaton mass term positive,  $M_\phi^2 >0$. 

The Lagrangian for the scale- and chirally invariant 
action at leading order ${\cal O}(p^2)$ is thus constructed from terms in Eqs.(\ref{inv:part}), (\ref{hard:part}) 
and (\ref{soft:part}): 
\begin{eqnarray} 
{\cal L}_{(2)} = {\cal L}_{(2)}^{\rm inv} + {\cal L}^{S}_{(2){\rm hard}} + {\cal L}^S_{(2){\rm soft}}  
\,. 
\label{L:p2} 
\end{eqnarray}   
From this we finally read off the dilaton mass term $\phi^2$ as~\cite{footnote3}
\begin{eqnarray} 
  M_\phi^2 
 =  m_\phi^2
+ (1+\gamma_m) (3-\gamma_m) \frac{N_f  F_\pi^2 m_\pi^2}{2F_\phi^2}\,.   
\label{SA:mf:nonzero}
\end{eqnarray}
Our result can also be derived directly from the underlying gauge theory through Eq.(\ref{WI}) 
as~\cite{footnotethree}:
\begin{eqnarray} 
\langle 0| \theta_\mu^\mu|\phi \rangle 
&=& \langle 0|\frac{\beta_{\rm NP}(\alpha)}{4\alpha} G_{\mu\nu}^2 |\phi \rangle \\ \nonumber
&+& (1+\gamma_m) N_f m_f \langle 0| \bar{\psi}\psi |\phi \rangle 
\,. 
\label{directderivation}
\end{eqnarray}  
We may further rewrite  
the dilaton mass, Eq. (\ref{SA:mf:nonzero}), in a form convenient for 
lattice simulations:
 \begin{eqnarray} 
  M_\phi^2   
&=&  m_\phi^2 + 
s\cdot m_\pi^2, 
\nonumber\\
s&\equiv& \frac{(3-\gamma_m)(1+\gamma_m)}{4}  
\cdot \frac{2N_f F_\pi^2}{F_\phi^2}  
\simeq \frac{2N_f F_\pi^2}{F_\phi^2} \equiv r
, \label{mphi:p2}
 \end{eqnarray}
where
the prefactor  $(3-\gamma_m)(1+\gamma_m)/4 = 1 - (\delta/2)^2 \simeq 1$ 
($\delta \equiv 1-\gamma_m$;  $(\delta/2)^2 \ll 1$) is very insensitive to the exact value of $\gamma_m$ as long  as  $\gamma_m \simeq 1$ in walking gauge theory.

\begin{figure}[htbp]
\begin{center}
   \includegraphics[scale=0.40]{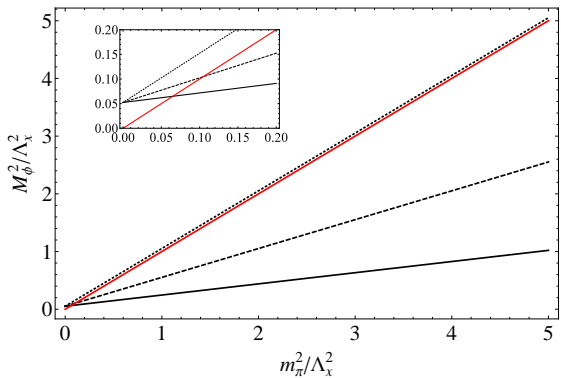}
\caption{ 
A plot of $M_\phi^2/\Lambda_\chi^2$ with respect to $m_\pi^2/\Lambda_\chi^2 (\equiv {\cal X} )$ obtained from Eq.(\ref{mphi:p2}), 
with $N_f= 8$ and $F_\pi=123$ GeV and the chiral-limit dilaton mass $m_\phi=$ 125 GeV. 
The slope $s\simeq r =2N_f F_\pi^2/F_\phi^2$ in Eq.(\ref{mphi:p2}) has been taken to be 0.2 (solid black), 0.5 (dashed black) and 1.0 (dotted black). 
The solid  red line corresponds to $M_\phi^2=m_\pi^2$. 
\label{Leading-TDMass}}
\end{center} 
 \end{figure}

  This is our main result.  It is useful for {\it determining simultaneously the chiral limit values of both the mass $m_\phi$ and the decay constant $F_\phi$}
 of the flavor-singlet scalar meson as the technidilaton of 
 walking technicolor on the lattice. Simultaneously fitting  
 the intercept and the slope of 
 a plot of $M_\phi^2$ vs $m_\pi^2$ from  
 the lattice data would give 
 $m_\phi^2$ (intercept) and 
 $F_\phi$ through the slope parameter $s \simeq r \equiv \frac{2N_f F_\pi^2}{F_\phi^2}$~\cite{Nf-dependence}.  
For a given $N_f$ all the quantities $\gamma_m$, $F_\pi$, $F_\phi$ and $m_\pi$ in the expression of 
the slope parameter $s$ 
can be measured separately in lattice simulations on the same set up.  
Hence {\it measuring  $s$ would be  a  self-consistency check 
of the simulations as a dilaton observation}, when compared with the value of $F_\phi$ determined by some other way.
In Fig.~\ref{Leading-TDMass}  we present plots $(x,y)=(m_\pi^2,M_\phi^2)$ of mock-up data for general case $s\simeq r=(0.2, 0.5, 1.0)$
in the one-family model, $N_f=8$ (4 weak-doublets) with $F_\pi=v_{\rm EW}/\sqrt{4}\simeq 123$ GeV,
by normalizing the masses to a chiral breaking scale $\Lambda_\chi = 4\pi F_\pi/\sqrt{N_f}$. 
The first number ($s=0.2$) corresponds to  
a phenomenologically favorable value~\cite{Matsuzaki:2012mk,Matsuzaki:2012xx}, 
$F_\phi \simeq \sqrt{2N_f} F_\pi/0.44 \simeq 1.1$ TeV, 
consistent with the current Higgs boson data 
at LHC. 
The third one ($s=1.0$) is the holographic estimate in the large $N_c$ limit~\cite{Matsuzaki:2012xx}. 
The  second value $(s=0.5)$ is just a sample number in between. 
The close-up window on the top-left panel in the figure shows that the dilaton mass 
gets larger than $m_\pi$ when the ChPT expansion parameter ${\cal X}\equiv m_\pi^2/\Lambda_\chi^2 =N_f m_\pi^2/(4\pi F_\pi)^2 \lesssim 0.06(0.1)$ for $s=0.2(0.5)$.  
Note also that for $s<1$ {\it there exists a crossing point} 
where $M_\phi^2<m_\pi^2$ changes to $M_\phi^2 >m_\pi^2$ near the chiral limit, as noted in Ref.~\cite{Aoki:2013qxa}.

As in the case of the usual ChPT~\cite{Gasser:1983yg}, 
chiral logarithmic corrections at the loop level 
would modify the chiral scaling 
of the dilaton mass formula in Eq.(\ref{mphi:p2}). 
Since the dilaton remains 
massive  
in the chiral limit due to the nonperturbative scale anomaly, 
only the pion loop corrections 
become significant for the chiral scaling of the dilaton mass. 
 Such chiral logarithmic corrections will be operative in the soft-pion region $m_\pi \lesssim M_\phi$   
(corresponding to the region where ChPT is valid: ${\cal X}\equiv m_\pi^2/\Lambda_\chi^2 \lesssim 0.1$ in Fig.~\ref{Leading-TDMass}).  
We shall compute the chiral logarithmic corrections coming from the pion loops 
arising from the vertices at the leading ${\cal O}(p^2)$ Lagrangian Eq.(\ref{L:p2}).  
Those corrections softly break the scale symmetry by the form $\sim (1,r) \cdot {\cal X} 
\log{\cal X}$ when the cutoff $\Lambda$ is identified with $\Lambda_\chi$,  
which will be renormalized by the soft-breaking ${\cal O}(p^4)$ counterterms proportional to $m_\pi^2 \sim {\cal M}$.

Using 
dimensional regularization~\cite{regularization}  
we thus find the $D=4$ pole (logarithmically divergent) contributions to the terms in quadratic order of dilaton fields: 
\begin{eqnarray} 
 \frac{1}{2} Z_{F_\phi} \partial_\mu \phi \partial^\mu \phi - 
\frac{1}{2} \tilde{m}_\phi^2 \phi^2 
\,, \label{L:p2p4}
\end{eqnarray} 
where 
\begin{eqnarray} 
Z_{F_\phi} &=& 1 + r\cdot \frac{N_f^2 -1}{2 N_f^2
} {\cal X} 
\log \frac{\Lambda^2}{m_\pi^2} 
\,,\nonumber  \\ 
\tilde{m}_\phi^2 
&=&
\Bigg[ 
m_\phi^2  - r m_\pi^2 \cdot \frac{2(N_f^2 -1)}{N_f^2 
}  {\cal X} 
\log \frac{\Lambda^2}{m_\pi^2} 
\,.   \nonumber 
\nonumber \\ 
&& 
+ \frac{(3-\gamma_m)(1+\gamma_m)}{4}\cdot  r m_\pi^2 \cdot
Z_{F_\phi} Z_{F_\pi}^{-1} Z_{m_\pi}^{-1}
 \Bigg] 
\,,\nonumber 
\end{eqnarray} 
with 
$
Z_{(i=F_\pi,m_\pi)} =  1 + \Gamma_i/N_f  \cdot {\cal X} 
 \log(\Lambda^2/m_\pi^2) 
$ 
and 
$\Gamma_{F_\pi}=N_f/4$ and $\Gamma_{m_\pi}=-1/N_f$. 
  After renormalizing the divergent parts at the renormalization scale $\mu$~\cite{counterterm}
 and defining the renormalized dilaton field 
$\phi_r = \sqrt{Z_{F_\phi}} \phi$, we find the renormalized $\phi^2$ terms,  
$ \frac{1}{2} \partial_\mu \phi_r \partial^\mu \phi_r - \frac{1}{2} M_\phi^2 \phi_r^2 $, 
with the dilaton mass including the chiral logarithmic corrections of ${\cal O}(p^4)$: 
\begin{eqnarray}
M_\phi^2 
&=& 
m_\phi^2 \left[ 1 +  r\cdot \frac{N_f^2-1}{2N_f^2}  
{\cal X}
 \log \frac{m_\pi^2}{\mu^2} \right] 
\nonumber \\ 
&& 
+  r \cdot m_\pi^2 \left[ \frac{  2(N_f^2 -1)}{N_f^2}  
{\cal X} 
\log \frac{m_\pi^2}{\mu^2} \right]
\nonumber \\ 
&& + 
s \cdot m_\pi^2  
\left[ 1 + \frac{ N_f^2 -4 }{4 N_f^2} 
{\cal X}
\log \frac{m_\pi^2}{\mu^2}  \right]
\nonumber\\ 
&& 
+ \left( \textrm{counterterms renormalized at $\mu$} \right) 
\,.  \label{mphi:ChiralLog}
\end{eqnarray}

  \begin{figure}[htbp]
\begin{center}
   \includegraphics[scale=0.35]{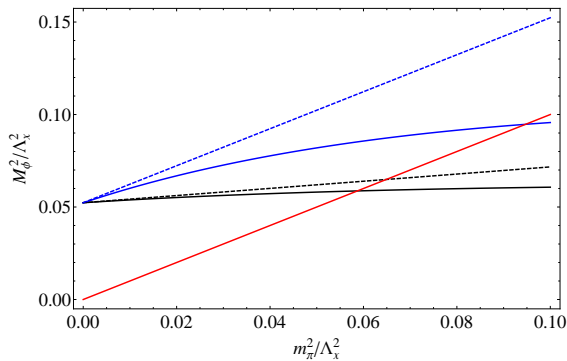}
\caption{ 
A plot of $M_\phi^2/\Lambda_\chi^2$ with respect to $m_\pi^2/\Lambda_\chi^2 (\equiv {\cal X} )$ including 
the chiral logarithmic corrections in Eq.(\ref{mphi:ChiralLog})  
for the one-family model with $N_f=8$, $F_\pi=123$ GeV, the chiral-limit mass $m_\phi=125$ GeV and 
the factor $s\simeq r =2N_f F_\pi^2/F_\phi^2=(0.2$, 1.0) (solid black and blue curves). 
 The leading-order scalings in Eq.(\ref{mphi:p2}) 
are also depicted for $s=0.2$ and 1.0 
by dashed black and blue lines, respectively. 
The solid red line corresponds to $M_\phi^2=m_\pi^2$. 
\label{ChiralLogEff-TDMass}}
\end{center} 
 \end{figure}  
 
We may assume that all the counterterms in Eq.(\ref{mphi:ChiralLog}) 
vanish at $\mu=\Lambda_{\chi}$, so that they  are induced only 
by the pion loops in the sChPT.  
As a concrete example, we again consider the one-family model with $N_f=8$ and $F_\pi=123$ GeV,  
and take the factor $s\simeq r =2 N_f F_\pi^2/F_\phi^2=0.2, 1.0$ and the chiral-limit dilaton mass $m_\phi=125$ GeV 
in the light of the LHC. 
In Fig.~\ref{ChiralLogEff-TDMass} we  
plot the chiral scaling behavior of the dilaton mass for a small pion mass region ${\cal X} \equiv m_\pi^2/\Lambda_\chi^2 \lesssim 0.1$, 
including the chiral logarithmic  
corrections from the pions at the one-loop level for $s=0.2$ and 1.0 (solid black and blue curves). 
 Also plotted is the leading-order formula  
 in Eq.(\ref{mphi:p2}) (dashed black and blue curves). 
The figure implies that the chiral logarithmic effect may be appreciable  
for the soft-pion mass region. 
However, such chiral logarithmic effects are negligibly small for the current  status of  $N_f=8$ QCD on the lattice, 
where  
simulations have been performed for a larger pion mass region $3 \lesssim {\cal X} \lesssim 5$~\cite{Aoki:2013xza}.

In conclusion,
 we have established a scale-invariant chiral perturbation theory (sChPT) for the pseudo-NG bosons, the pion ($\pi$) and the dilaton ($\phi$),
which will be useful in its own right in various situations. 
It is straightforward~\cite{Kurachi:2014qma} to include the vector mesons into this framework via hidden local 
symmetry \cite{Bando:1987br}.  
As its prominent consequence we obtained a formula relating the masses $M_\phi^2$ vs $m_\pi^2$, Eq.(\ref{mphi:p2}) (tree), or Eq.(\ref{mphi:ChiralLog})
(one-loop), which we believe plays a vital role for making chiral extrapolations of 
lattice data of  
the flavor-singlet scalar meson, thereby obtaining the mass ($m_\phi$) and decay constant ($F_\phi$) of 
the technidilaton as a composite Higgs boson in walking technicolor.

We would like to express our sincere thanks to all the members of LatKMI Collaboration for helpful discussions and information.
This work was supported in part 
by the JSPS Grant-in-Aid for Scientific Research (S) \#22224003 and (C) \#23540300 (K.Y.). 

 [Note added]
 After submission of  this article, the LatKMI Collaboration published a paper \cite{Aoki:2014oha} (follow-up of \cite{Aoki:2013qxa} ) 
 finding a light flavor-singlet scalar in $N_f=8$ QCD, 
 with the data analyzed based on Eq. (\ref{mphi:p2})
 to be roughly consistent with 125 GeV Higgs as the technidilaton.


\begin{thebibliography}{9}
\bibitem{footnoteone} 
The Lagrangian is constructed uniquely (up to total derivatives) by the requirement that the action $S[\varphi(x)]$ $=$ $ \int d^4x \, {\cal L}[\varphi(x)]$
be invariant under the scale (and also chiral) transformation 
$\delta \varphi(x)$ $=$ $(d_\varphi + x_\mu \partial^\mu)\varphi(x)$, $\delta x_\mu=- x_\mu$. 
This implies that $\delta S[\varphi(x)] = \int ( \delta d^4 x\, {\cal L} + d^4x \,\delta{\cal L} ) 
=\int d^4 x (- 4 {\cal L} + d_{\cal L} {\cal L})=0$, 
namely the scale dimension of 
${\cal L}$ must be 4,
$d_{\cal L}$ $=$ $4$. Eq.~(\ref{inv:part}) is a unique chirally invariant Lagrangian having 
scale dimension four  
and a correct kinetic term of $\varphi$ $=$ $(\phi,\pi)$. 
Note that the dimensionless field $\chi(x)$ transforms as an operator with the scale dimension one, with $\langle \chi \rangle$ $=$ $1$.   
Similarly, the Lagrangians, Eqs.~(\ref{inv:part}), (\ref{hard:part}), and (\ref{soft:part}) 
are unique in a way to have the scale-dimension four and to satisfy  
the anomalous WT identities 
as well as the stable vacuum. 
See also \cite{footnote2}.
\bibitem{addfootnote}
The scale-WT identity for the operator ${\cal O}$ reads 
$0$ $=$ $\lim_{q\rightarrow 0}\, q_\mu \int d^4 x e^{- i q x} \langle 0| T(D^\mu(x) \, {\cal O} (0)) |0 \rangle $ 
$=$ $\langle 0 | [i Q_D,  {\cal O} ]| 0\rangle$ 
$+$ $  i\int d^4 x \langle 0| T(\partial_\mu D^\mu(x)\,  {\cal O} (0) ) |0 \rangle $
$=$ $ d_{\cal O} \, \langle {\cal O} \rangle - F_\phi \cdot \langle \phi|{\cal O} |0 \rangle$, 
where $[i Q_D,{\cal O}] =d_{\cal O} \, {\cal O}$, with  $Q_D= \int d^3 x D^0(x)$,  and $d_{\cal O}$ 
is the scale dimension of the operator ${\cal O}$. Note that 
the second term comes from 
the $\phi$ pole contribution.  
Thus we have $d_{\cal O} \langle 0| {\cal O} |0 \rangle = F_\phi \cdot \langle 0| {\cal O}  |\phi\rangle$.
\bibitem{footnotethree}
Note that $\langle 0| \theta_\mu^\mu|\phi \rangle
_{m_f \neq 0}$ $=$ $\langle 0| 
\partial_\mu D^\mu |\phi \rangle
_{m_f \neq 0}$ $=$ $ - F_\phi M_\phi^2$, 
$\langle 0|\frac{\beta_{\rm NP}(\alpha)}{4\alpha}  G_{\mu\nu}^2|\phi \rangle$ $=$ $ \langle 0| 
\partial_\mu D^\mu |\phi \rangle
_{m_f = 0} $ $=$ $ - F_\phi m_\phi^2$, 
while the second term on the  RHS of Eq.(\ref{SA:mf:nonzero}) comes from the scale-WT identity (see \cite{addfootnote}): 
$\langle 0| m_f \bar{\psi}\psi |\phi \rangle$ $=$ $(3-\gamma_m)$ $2 m_f \langle \bar{\psi}\psi \rangle/(2 F_\phi) $, when
combined with the chiral-WT identity (Gell-Mann-Oakes-Renner relation): 
$  2 m_f \langle \bar{\psi}\psi \rangle\delta^{\alpha \beta}$ $=$ $\langle [i Q^\beta_5, 2 m_f \bar \psi i \gamma_5 T^\alpha \psi] \rangle $ 
$=$ $F_\pi \langle 0| \partial_\mu J_5^{\alpha,\mu} |\pi^\beta \rangle$ $=$ $- F^2_\pi m_\pi^2 \delta^{\alpha\beta}$, 
where $Q_5^\alpha = \int d^3 x J_5^{\alpha,0}(x)$. 
 (Note that $\theta_\mu^\mu$  in Eq.(\ref{WI}) need not be a subtracted one in this derivation, since the constant subtraction $\langle \theta_\mu^\mu \rangle_{\rm perturbation}$
sandwiched by $\langle 0|$ and $|\phi\rangle$ is zero.)
\bibitem{Nf-dependence} 
Note that $r$ has no explicit $N_f$-independence, since $F_\phi^2$ is associated with the flavor-singlet operator having a sum of 
$N_f$-flavor contributions, namely $F_\phi^2$ is proportional to $N_f$, while $F_\pi^2$ is not.
\end{thebibliography}
\end{document}